\documentclass[12pt]{article}
\usepackage{graphicx}
\begin{document} 

\title{Advertising, consensus, and ageing in multilayer Sznajd model}

\author{Christian Schulze\\
Institute for Theoretical Physics, Cologne University\\D-50923 K\"oln, Euroland}

\maketitle
\centerline{e-mail: ab127@uni-koeln.de}

\bigskip
Abstract:  
In the Sznajd consensus model on the square lattice, 
two people who agree in their
opinions convince their neighbours of this opinion. We generalize
it to many layers representing many age levels, and check if
still a consensus among all layers is possible. Advertising
sometimes but not always produces a consensus on the advertised
opinion.

\bigskip


\begin{figure}[hbt]
\begin{center}
\includegraphics[angle=-90,scale=0.5]{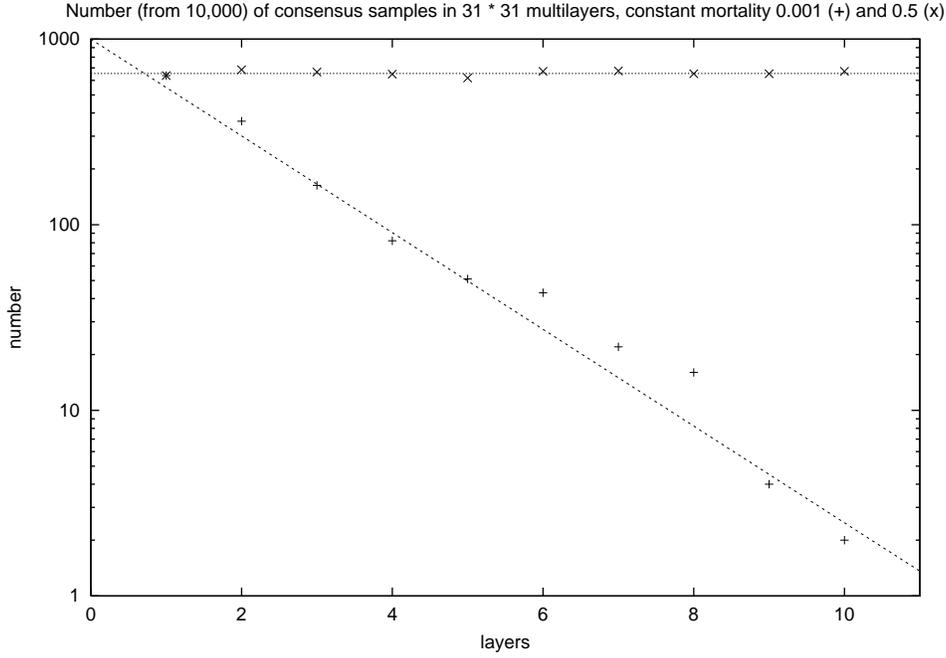}
\end{center}
\caption{ 
Number of samples, from $10^4$, where for $m=4$ a consensus is reached, versus
the number $N$ of layers or age levels. For
high mortalities the consensus probability stays at about 7 \% while at low
mortalities it decays exponentially with $N$, nearly as $1/4^N$.
}
\end{figure}

\begin{figure}[hbt]
\begin{center}
\includegraphics[angle=-90,scale=0.5]{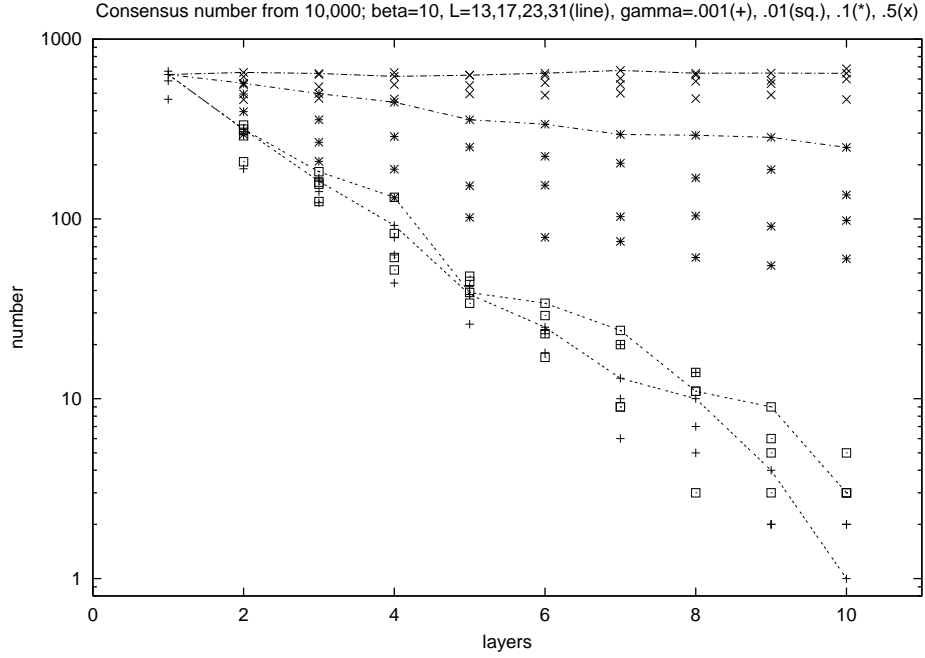}
\end{center}
\caption{ 
As Fig.1 but with ageing, for four mortality factors $\gamma$ = $1/10^3, \;
1/10^2, \; 1/10, \; 1/2$ (from bottom to top) and four different lattice sizes
$L = 13, \, 17, \, 23, \, 31$. The largest lattices are symbolized by lines.
}
\end{figure}

\begin{figure}[hbt]
\begin{center}
\includegraphics[angle=-90,scale=0.5]{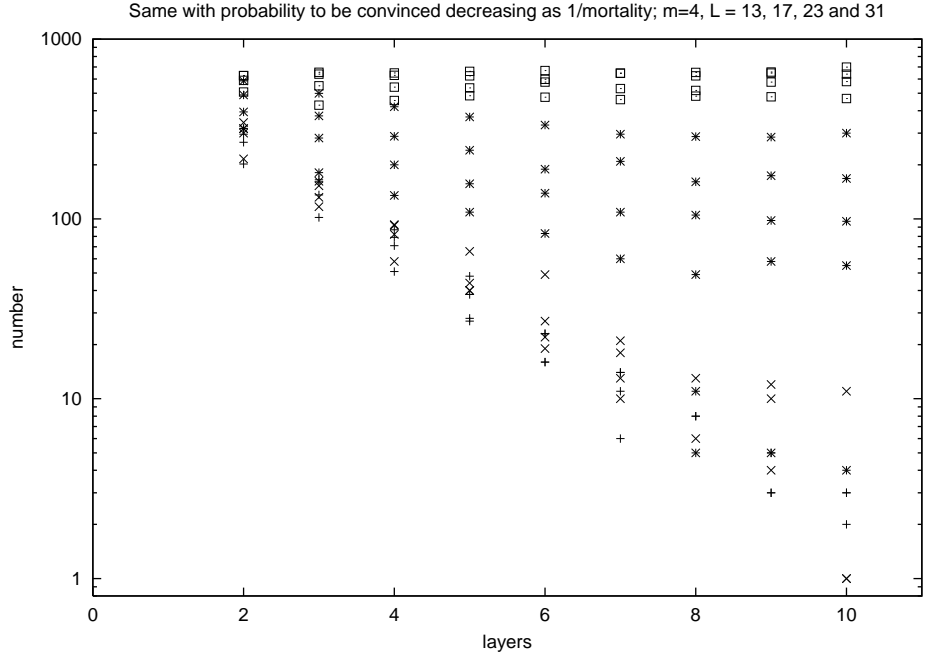}
\end{center}
\caption{ 
As Figs.1,2 but with convincing probability decreasing exponentially with age.
}
\end{figure}

\begin{figure}[hbt]
\begin{center}
\includegraphics[angle=-90,scale=0.5]{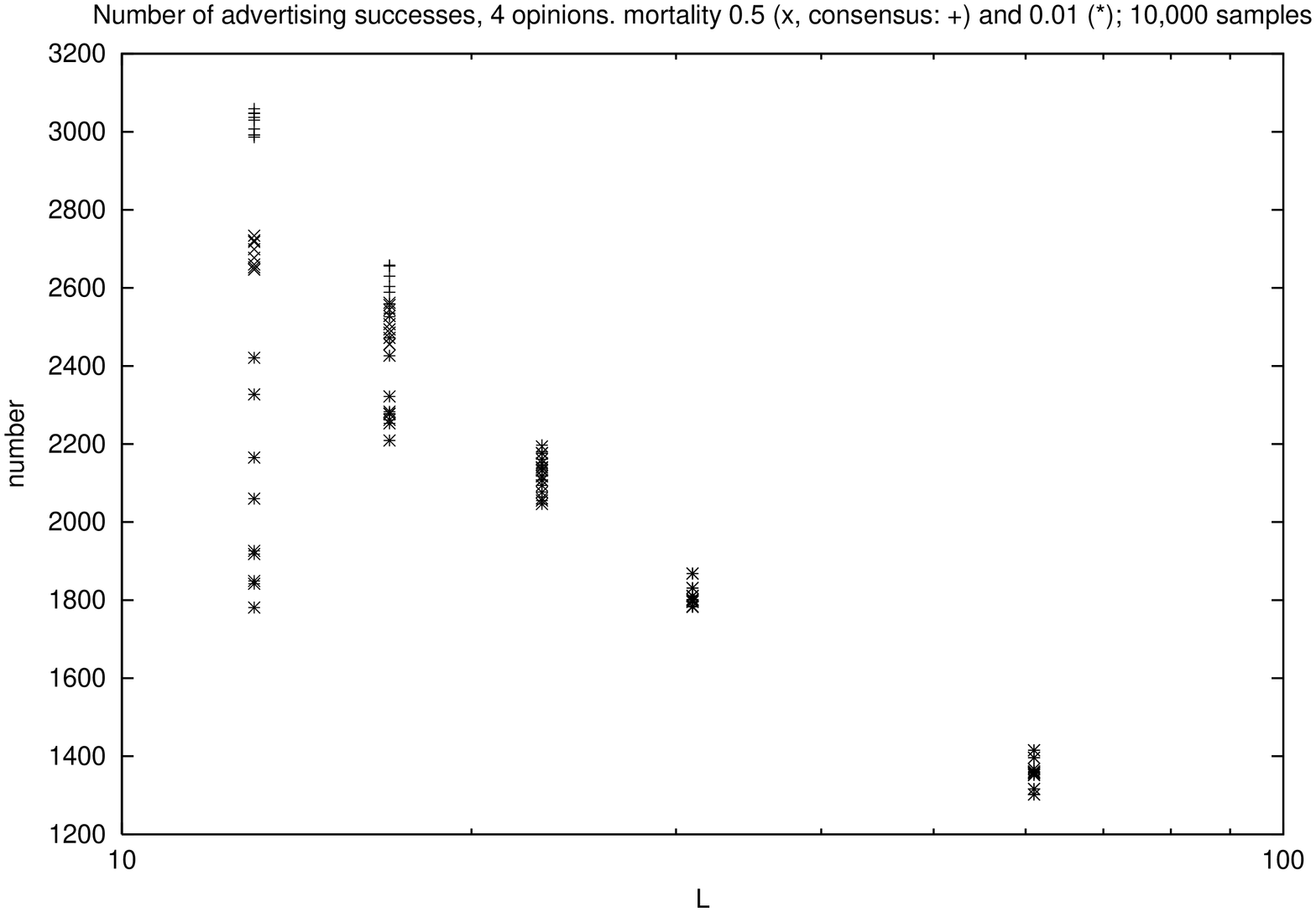}
\end{center}
\caption{ 
Number, among $10^4$ samples, of consensus (+) and of success (x,*) cases, for
constant low (*) and high (+,x) mortality and $m=4$. A consensus is called
a success if the consensus opinion is the advertised one. For low mortality
nearly every consensus is a success and thus the consensus numbers are not
shown.
}
\end{figure}

\bigskip

Many recent models tried to simulate how in a society a consensus
may emerge \cite{axelrod,weisbuch,hegselmann,redner}. Particularly
well studied is the Sznajd model \cite{sznajd,stauffer}
where two neighbouring people (on a one-dimensional chain or 
higher-dimensional lattice) convince their neighbours if and
only if these two people agree in their opinions. We generalize
here the square lattice to a multilayer model, where each layer 
correponds to a different age group of the simulated people.
We check if the phase transition known for the simple square 
lattice is also valid in the multilayer model and ask how many
different initial opinions are allowed if a consensus is still 
desired. Finally, we check the effects of advertising 
\cite{schulze,sznajd2} favouring the first of the several 
possible opinions.

Each site on an $L \times L$ square lattice has one of $m$ 
possible opinions, initially random. A pair of nearest neighbours, randomly 
selected in sequential updating, may share the same opinion $j$.
In this case, all six neighbours of this pair are convinced of
that opinion, that means they are forced to adopt the same
value $j$. This model has a phase transition \cite{stauffer}
as a function of the initial concentration $p$: If for $m=2$
the initial distribution of opinions is random and if the 
fraction $p$ of sites has the first and the remaining fraction
$1-p$ the other opinion, then for large lattices everybody ends up
with the first opinion if $p < 1/2$ and everybody with the other
opinion if $p > 1/2$. If $m > 2$ we assume bounded confidence 
\cite{hegselmann,weisbuch}, which means that no neighbours 
can be convinced which differ by more than one opinion unit from the
central pair. In that case \cite{stauffer} $m=3$ usually still leads to 
a consensus while $m=4$ seldomly allows a consensus: It is difficult to
form a stable government coalition with four and more parties.

Now we generalize this planar Sznajd model to a multilayer model of $N$
square lattices on top of each other, representing $N$ age intervals in
the life of an individual. With a certain probability $q$, called mortality,
an individual dies, the younger ones on the same two-dimensional site all
move up one time unit in their age, and a baby is born on the newly freed site,
having the opinion of the parent. Within each layer only, and not between 
layers, the usual Sznajd convincing process takes place. 
First, we found that for two layers we still
have a phase transition. Below we will check the chances of finding a consensus
as a function of the number $N$ of layers in four cases: with an age-independent
mortality (Fig.1), with a mortality increasing exponentially with age (Fig.2), 
with a probability to be convinced inversely proportional to mortality (Fig.3),
and finally again with age-independent mortality and convincing probability but
a constant advertising effort in favour of the first of $m$ opinions (Fig.4).

In this multilayer model we continue the simulations until the basic layer
(the babies) reaches a complete consensus within that layer only. Then we 
find the majority opinion in each of the higher layers, and we declare
a consensus if and only if the majority opinion in each of the higher layers 
agrees with the one surviving opinion in the basic layer.
  
With $m \le 3$ possible opinions, a consensus is always formed for large
lattices and long enough times, while for $m \ge 4$ a consensus is rare, just
as for the single layer. We used $L = 13, 17, 23, 31, 61$ and sometimes 101. 
Fig.1 shows for a $31 \times 31$ lattice that for
small mortalities $q = 0.001$ the probability to reach a consensus, already 
below 7 percent for a single layer, decays exponentially towards zero with
increasing number $N$ of layers. In contrast, for large mortalities, $q=0.5$,
it stays at its single-layer value for all $2 \le N \le 10$. The higher 
mortality leads to a stronger coupling between layers and thus to easier 
overall consensus; note that no convincing takes place between different layers.

This trend of Fig.1 is well reproduced in Fig.2 if we include ageing, that means
a mortality $q$ increasing exponentially with age $n = 1,2, \dots, N$ according
to the Gompertz law of the 19th century: $q=\gamma \exp[10(n-N)/(N-1)]$. Not 
much changes if we include that young people are more easily convinced than 
old people and thus take a convincing probability $\exp[10(N-n)/(N-1)]=\gamma/q$
in Fig.3.

Advertising has the effect that independent of the whole convincing process,
a site assumes the first opinion with probability 0.1, at every iteration. The
mortality is taken again as 0.01 independent of age, and convincing always
takes place. For $m=2$ opinions, advertising always wins for large enough
lattices, just as in a single layer \cite{schulze,sznajd2}; also for $m=3$
advertising reaches a complete consensus for the advertised opinion. Fig.4
shows a more complicated behaviour for $m=4$: Now with increasing lattice size
the fraction of advertising successes shrinks, possibly towards zero. It is 
quite independent of the number of layers which varies from 2 to 10. Fig.4 for
small lattices like $L = 13$ also shows that sometimes a consensus is reached
but not for the advertised opinion.

In summary, the multilayer model is similar to the single-layer Sznajd model in
allowing a consensus for two and three opinions but not for four (and more).
This effect remains if advertising, a mortality increasing with age, and a 
convincing probability decreasing with age are introduced. The low probability
of consensus, in the case of four opinions without advertising, decreases 
exponentially with the number of age layers. 

Thanks are due to D. Stauffer for help.


\end{document}